\documentclass[prd,aps,preprint,tightenlines,superscriptaddress]{revtex4}
\usepackage{epsfig}
\usepackage{color}
\usepackage{amsmath,slashed}
\begin{document}


\title{Proton Tomography Through Deeply Virtual Compton Scattering}
\author{Xiangdong Ji}\thanks{The 2016 Feshbach prize in theoretical nuclear physics talk
given at APS April Meeting, Salt Lake City, Session U3: DNP Awards Session, Monday, April 18, 2016. The write-up is slightly expanded in details
from the original talk.}
\affiliation{Maryland Center for Fundamental Physics, Department of Physics, University of Maryland, College
Park, Maryland 20742, USA}
\affiliation{INPAC, Department of Physics and Astronomy, Shanghai Jiao Tong University, Shanghai, 200240, P. R. China}

\date{\today}
\vspace{0.5in}
\begin{abstract}

In this prize talk, I recall some of the history surrounding the discovery of deeply virtual Compton scattering, and explain why
it is an exciting experimental tool to obtain novel tomographic pictures of the nucleons at Jefferson Lab 12 GeV facility
and the planned Electron-Ion Collider in the United States.

\end{abstract}

\maketitle
It is certainly a great honor to have received the 2016 Herman Feshbach Prize in theoretical nuclear physics by the American Physical Society (APS).
I sincerely thank my colleagues in the Division of Nuclear Physics (DNP) to recognize the importance
of some of the theoretical works I have done in the past, particularly their relevance to the
experimental programs around the world.

\section{Herman Feshbach and Me}

Since this prize is in honor and memory of a great nuclear theorist, Herman Feshbach, it is fitting for me to start my talk by recalling
some of my personal interactions with Herman. I first heard Feshbach back in 1982 when I was a freshman graduate student at
Peking University before I knew anything about nuclear physics. I saw my fellow roommate reading a bulky
book titled ``Theoretical Nuclear Physics: Nuclear Structure" by De Shalit and Feshbach~\cite{deshalit}. He told me it is the bible
in nuclear physics and every graduate student should read it. I was quite impressed and
thought, ``Well, someone who writes a bible must be a god!"

From 1989 to 1996, I spent 7 years at the Center for Theoretical Physics, MIT, first as a postdoc
and then as a junior faculty. Herman was around all that time, first as a mentor and then as a senior
colleague. He was humorous, generous and helpful as he told many interesting stories and
offered me many useful advices. In 1991, he finally finished
the sequel in Theoretical Nuclear Physics: Nuclear Reactions~\cite{feshbach}. I helped him proofread some of
the chapters. I remember asked him jokingly why he wanted to publish a book while no
one was going to buy it anymore. Well, I was clearly wrong. With the advent of the Facility for Rare Isotope Beams (FRIB) in the United States,
it becomes a great reference for many scientists and students. I am sure many copies will be sold (particularly in the state of Michigan!).

Around that time, Herman tried to convince me to work on heavy-ion physics. The Relativistic Heavy Ion Collider (RHIC) at Brookhaven National Laboratory just
started construction and it would generate a ``hot" field. I tried and even hired a postdoc who knows
thermal field theory to work on it. Unfortunately, I found that heavy-ion collisions are too complicated for my taste,
and quit half a year later.

\begin{figure}
\centering
\includegraphics[width=0.5\textwidth]{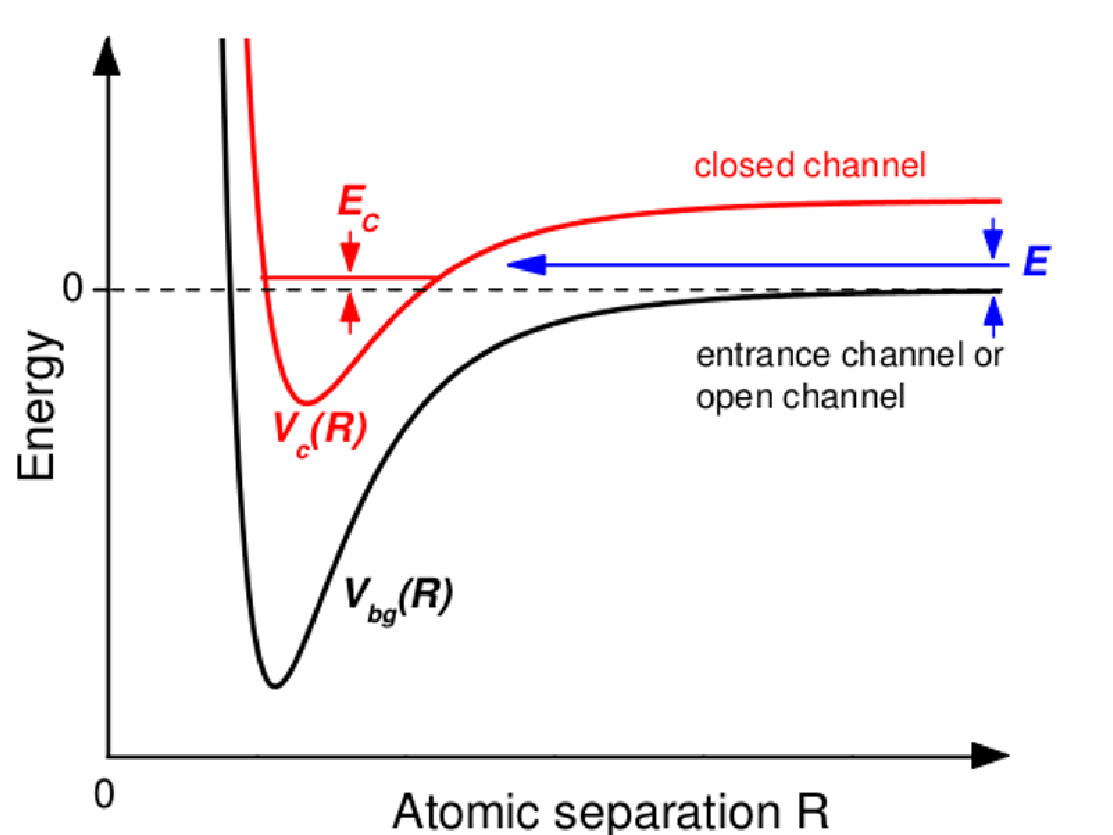}
\caption{Feshbach resonance in a two-body system with two channel couplings. Courtesy of Ravi Mohan.}
\label{resonance}
\end{figure}

\begin{figure}
\centering
\includegraphics[width=0.5\textwidth]{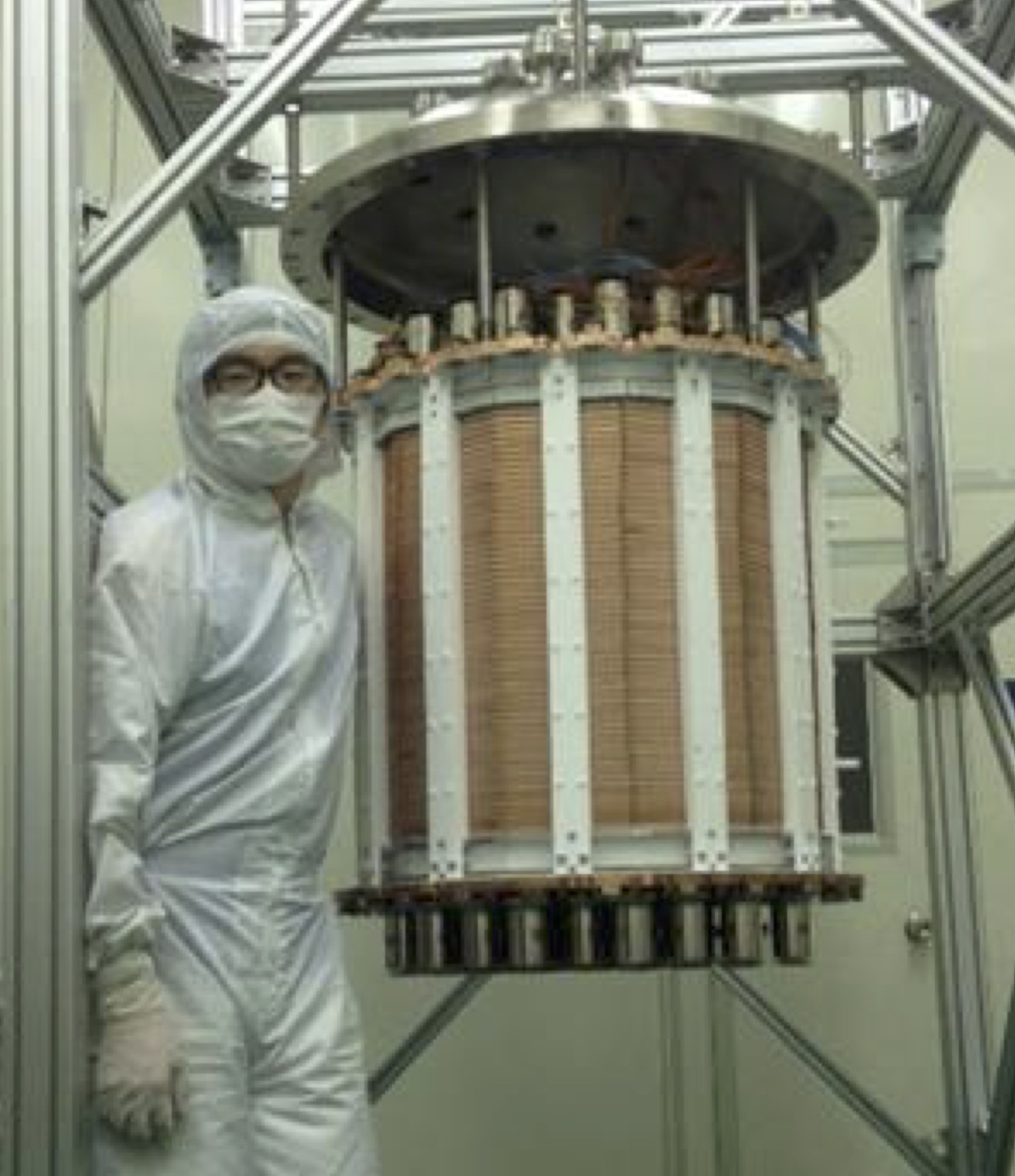}
\caption{The 500kg liquid Xe dark matter detector in the PandaX experiment, currently running in China Jinping Underground Lab.}
\label{pandax}
\end{figure}

In 2009, more than a decade after I left MIT, I started an interest in direct dark matter detection and
encountered for the first time Feshbach resonance. It is about two-channel coupling in an effective two-body
system. As shown in Fig.~\ref{resonance}, there may be a bound state in the closed channel. However, in the open channel, that bound state
energy corresponds to a scattering state.
The coupling between the two channels turns the bound state into a resonance. This is exactly the mechanism
of 178 nm photon production in liquid xenon (Xe). Two ground state Xe atoms cannot form a bound state
(the van der Waals attraction is too weak). However, when one of the Xe atoms is in the excited
state, the two will attract strongly enough to form a bound state. Since its energy is greater than zero
and the molecule can decay into two ground state Xe atoms by emitting a 178nm photon. In a liquid Xe
Time Projection Chamber, one uses special Photon Multiplier Tubes (PMT) to detect these photons that could be
generated by dark-matter particles direct scattering with a Xe nucleus. Shown in Fig.~\ref{pandax} is the 500kg liquid Xe
dark matter detector assembled by my graduate student Andi Tan, which shows the top and bottom layers
of 3" PMTs~\cite{Tan:2016diz}. Currently the detector is in operation in China Jinping Underground Lab, and is the largest
running dark matter detector of its kind in the world. Feshbach resonance makes this type of detection
possible, and understandable thanks to Herman's deep theoretical insight.

\section{Spin Crisis and Aftermath}

\begin{figure}
\centering
\includegraphics[width=0.8\textwidth]{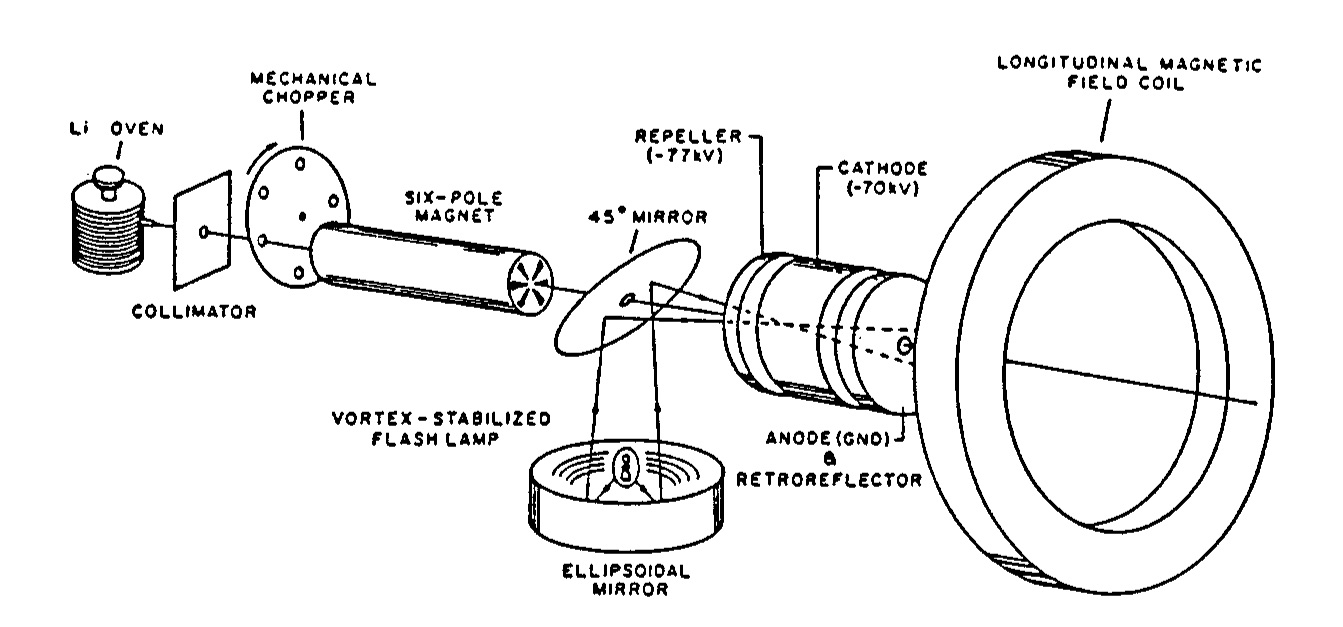}
\caption{Schematic diagram of the Yale polarized electron source, PEGGY, showing the principal components of the lithium atomic beam, the ultraviolet optics, and the ionization region electron optics~\cite{Hughes:2000ss}.}
\label{peggy}
\end{figure}

Now back to the main topic of my talk. This brings me to another great physicist of our time, Vernon Hughes.
According to NNDB~\cite{nndb}, Vernon studied under I. I. Rabi and investigated muons. Of course, he made seminal contributions
to muon physics~\cite{Hughes:2000ss} including the discovery of muonium~\cite{Hughes:1960zz} and the most precise measurement of muon anomalous magnetic moment $g-2$~\cite{Bennett:2006fi}.
However, many people may not know that Vernon also started the field of polarized deep-inelastic scattering (DIS).
His group at Yale was the first to produce the polarized electron source PEGGY (see Fig.~\ref{peggy}) that was brought to Stanford Linear Accelerator Center (SLAC). ``The first polarized electron beam was accelerated to 20 GeV in 1974, and a polarization of 0.8 was measured in M\o ller scattering"~\cite{Hughes:2000ss,Cooper:1975cu}. Vernon
can rightly be called ``the father of proton spin physics".

\begin{figure}
\centering
\includegraphics[width=0.6\textwidth]{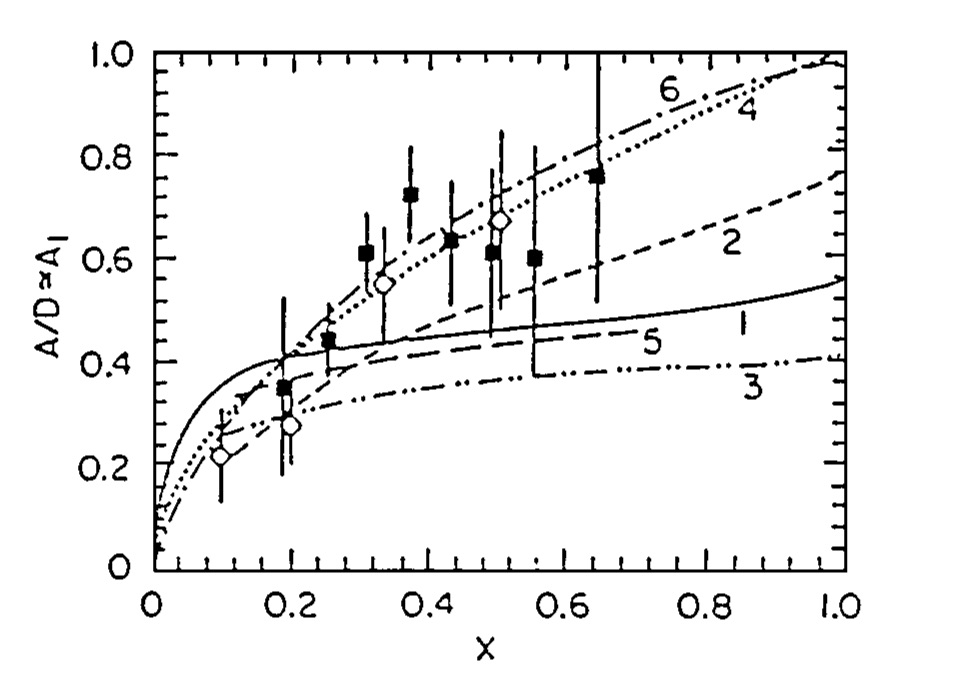}
\caption{Experimental values $A$ compared to different proton models. (1) Symmetric valence-quark model. (2) Current quarks. (3) Orbital angular momentum. (4) Unsymmetric model. (5) MIT bag model. (6) Source theory. For more detailed account see Refs.~\cite{Hughes:2000ss,Alguard:1978gf}.}
\label{asymmetry}
\end{figure}

In the 1970's, Vernon led the first generation of polarized DIS at SLAC, measured the spin asymmetry
\begin{equation}
A= \frac{d\sigma_{\uparrow\uparrow} - d\sigma_{\downarrow\uparrow}}{d\sigma_{\uparrow\uparrow} + d\sigma_{\downarrow\uparrow}}\ ,
\end{equation}
where $d\sigma_{\uparrow\uparrow}$ and $d\sigma_{\downarrow\uparrow}$ denote DIS cross sections when the spin of the electron is parallel and anti-parallel to that of the proton target. The early SLAC
data was shown in Fig.~\ref{asymmetry}, where comparison with various proton models were made~\cite{Alguard:1978gf}.

\begin{figure}
\centering
\includegraphics[width=0.7\textwidth]{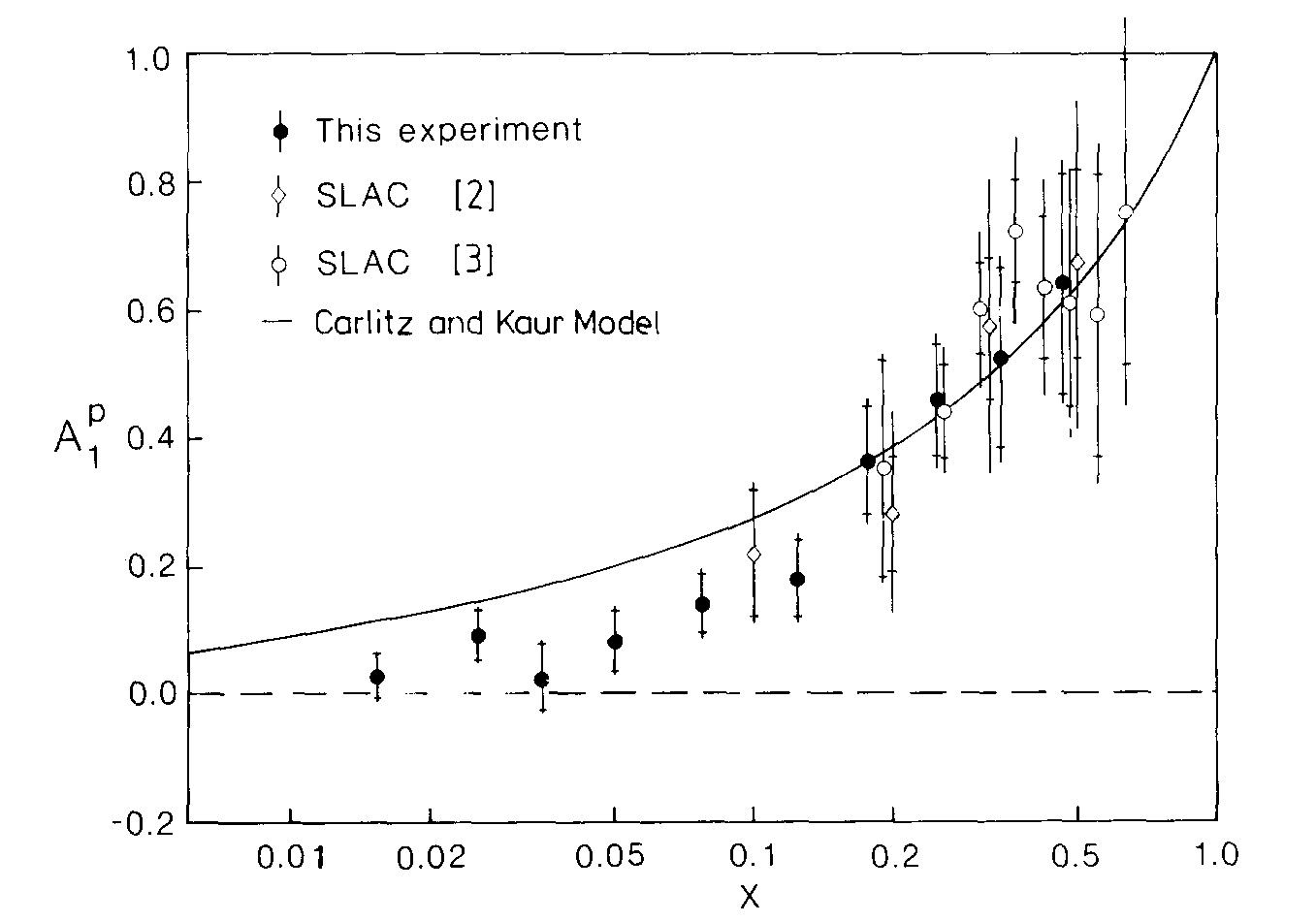}
\caption{The spin asymmetry $A$ for the proton as a function of $x$~\cite{Ashman:1989ig} compared with quark model prediction~\cite{Carlitz:1976in}.}
\label{emc}
\end{figure}

When SLAC turned down his request for further polarized DIS experiments in 1980's, Venon's group joined the European Muon Collaboration (EMC).
Together with Lancaster and Liverpool groups, he led the polarized DIS program at CERN. Their surprising result published in 1989
is shown in Fig.~\ref{emc}~\cite{Ashman:1987hv,Ashman:1989ig}, compared with a quark model calculation~\cite{Carlitz:1976in}. One clearly sees a systematic discrepancy in
the small-$x$ region. The asymmetry can be used to determine the fraction of the proton spin carried by quark spin.
The result is astonishingly small, $\langle S_z\rangle_{\text{quark}} = +0.06(8)$ with $S_{\text{proton}}=1/2$~\cite{Hughes:2000ss}. In Vernon's own words \cite{Hughes:2000ss}, ``This surprising result that the quark spins carry only a small fraction of the proton spin and, in addition, that the strange quarks have a negative polarization the spin crisis or, perhaps better, the spin puzzle."

The simple quark model cannot simply be dismissed~\cite{GellMann:1964nj}. It has been the most important clue
for the discovery of the fundamental theory of strong interactions, quantum chromodynamics (QCD). The quarks have spin
1/2 and the color degrees of freedom. These basic features have been inherited in the fundamental theory.
Many static properties of hadrons, including magnetic moments and weak decay constants,
are well predicted by the quark model. However, the model also predicts that the proton spin is entirely carried
by the spin of the unpaired quark.

In the following decade, a number of experiments worldwide were initiated to check the EMC spin result.
There were E145, E155 experiments led by Vernon's son, Emlyn Hughes. There was the Spin Muon Collaboration
(SMC) at CERN following the EMC. Then there was the HERMES experiment at DESY (Deutsches Elektronen-Synchrotron). There has been a lot
of theoretical work on the perturbative QCD corrections up to and beyond the next-to-leading order (NLO).
When the dust settles, we now know for a certainty that the quark spin accounts for about 1/3 of the proton
spin~\cite{Filippone:2001ux,deFlorian:2009vb}, a fraction far less than that predicted by the simple quark model.

The influence of the EMC result may be judged from the number of citations of the original two papers~\cite{Ashman:1987hv,Ashman:1989ig}. Both
the Physics Letter B and the Nuclear Physics B papers now have received more than 1500 citations according the Inspire
data base (https://inspirehep.net/).

It is interesting to note that historically the first evidence that the proton has a nontrivial structure
came from Otto Stern's measurement of the anomalous magnetic moment of the proton in Hamburg in 1933~\cite{stern1993}.
In 1934, G. Breit and I.~I. Rabi pulished a paper in Physical Review titled ``On the Interpretation
of the Present Values of the Nuclear Moments"~\cite{breit1934}. Rabi went to Europe at the end of the 1920's,
hoping to work with Pauli on theory. However, he ended up doing a spin-related experiment
at Stern's lab, where he launched his career in experimental molecular beam physics. Thus, Vernon's drive
for spin physics had a clear trail.

\vspace{.5cm}

So what did we learn from the spin crisis? On the theory side, like the Bohr model for the hydrogen atom,
the nucleon models built in the 60's and 70's had great historical importance, but now we have
the fundamental theory---QCD. We have to solve QCD directly
to understand the experimental result. Lattice QCD simulations, although still in the process of
reaching respectable precision for baryons (particularly, the computation of
the so-called disconnected diagrams), show that the theoretical expectation on the fraction of the nucleon spin carried
in quark spin is about 30\%~\cite{Deka:2013zha}. Thus there is no substantial discrepancy between
the fundamental theory and data.

On the experimental side, we are no longer satisfied with measuring just the charge and current distributions
through nucleon elastic form factors, and the quark distributions in Feynman momentum through DIS. We need to
know more. Two important directions have been pursued since early 1990's. One is the RHIC spin program~\cite{Bunce:2000uv}, and the other
follows the discovery of deeply virtual Compton scattering as a novel probe of the nucleon structure. The latter
is the main subject of this talk.

\begin{figure}
\centering
\includegraphics[width=\textwidth]{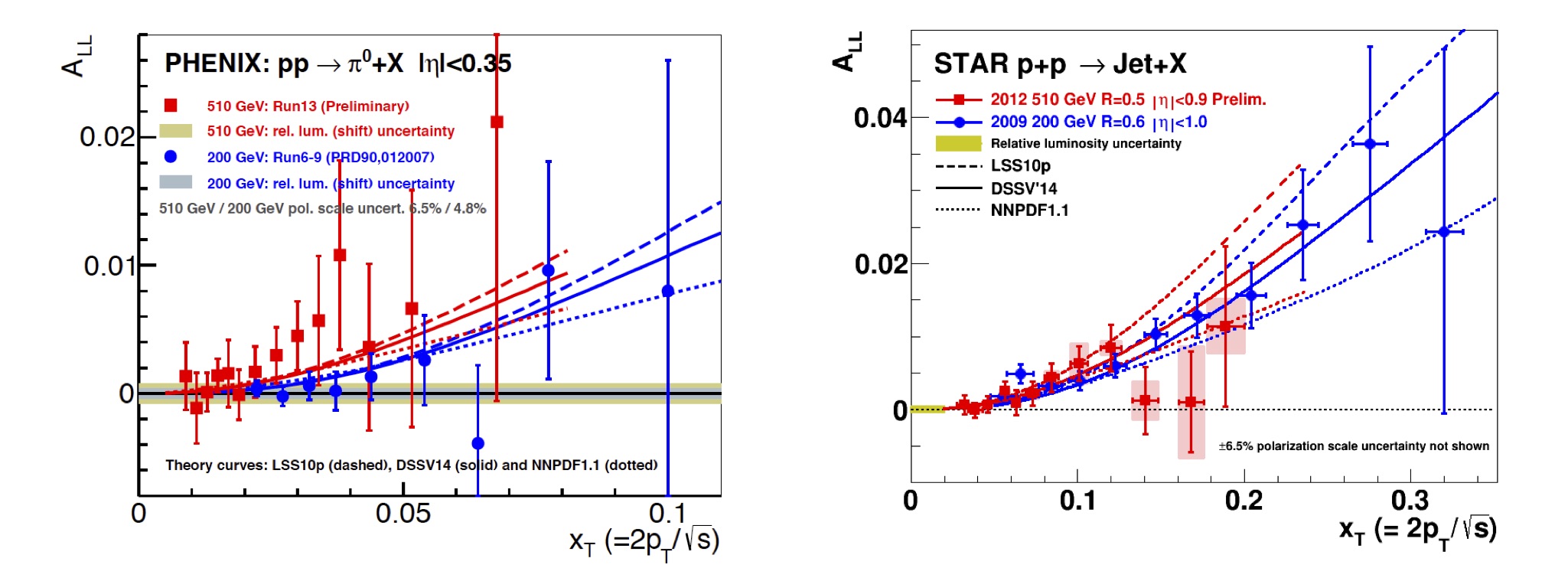}
\caption{The spin asymmetry for $\pi^0$-meson production (left) and jet production (right) at mid rapidity with point-to-point uncertainties in 200 GeV and 510 GeV $p+p$ collisions. The curves are recent NLO global analyses (for reference see Ref.~\cite{Aschenauer:2015eha}).}
\label{rhic}
\end{figure}

\begin{figure}
\centering
\includegraphics[width=0.6\textwidth]{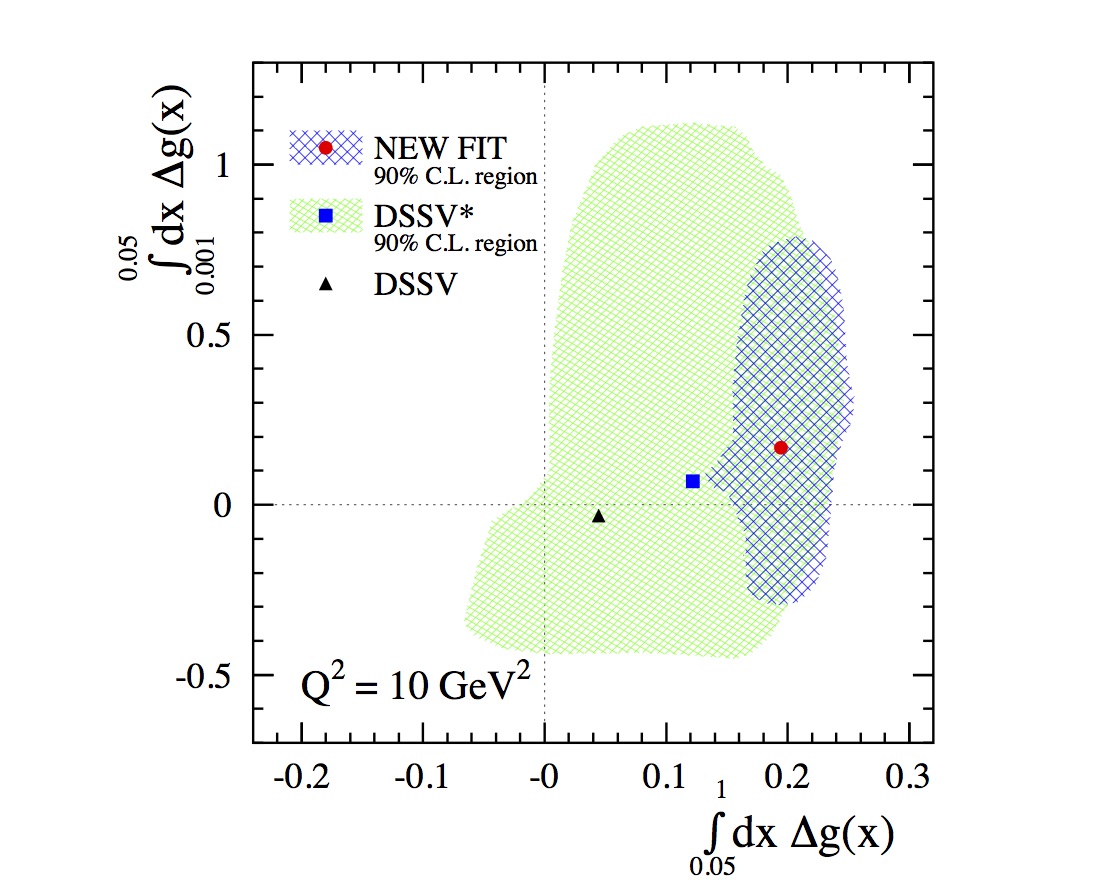}
\caption{90\% C.L. areas in the plan spanned by the truncated moments of $\Delta g(x)$ computed for $0.5\le x \le 1$ and $0.001\le x \le 0.05$ at $Q^2=10\text{GeV}^2$. The new analysis (red dot and blue shadow) based on RHIC data up to Run-2009 comes from Ref.~\cite{deFlorian:2014yva}.}
\label{deltaG}
\end{figure}

It is important to highlight a bit about the RHIC spin program and its success, which
was largely motivated by a once-popular theoretical argument that the gluon may have
significant polarization $\Delta G$, which cancel a large part of quark polarization through the axial anomaly~\cite{Altarelli:1988nr}. However,
the gluon contributes in a much more straighforward way to the proton spin in the infinite
momentum frame, as was put down by Jaffe and Manohar~\cite{Jaffe:1989jz},
\begin{equation}
                {1\over2} = {1\over2} \Delta \Sigma + \Delta G + l_q^z + l_g^z\ ,
                \label{jaffe}
\end{equation}
where $l_q^z$ and $l_g^z$ are canonical quark and gluon orbital contributions, respectively.
The idea of polarizing RHIC was certainly driven by the need to understand how big the gluon polarization there is in the proton. The generation
of polarized proton beams and completing the PHENIX detector with better spin physics capability have been
helped enormously by RIKEN in Japan. The RHIC spin program includes measurements of the gluon polarization
$\Delta G$ through $\pi^0$ and jet production, direct photon process and heavy flavor production, polarization
of valence and sea quarks through the single spin asymmetry in $W$ production, and transverse spin effects.
 The most recent result in Fig.~\ref{rhic} shows substantial amount of positive gluon polarization in the polarized proton,
perhaps at the level of $0.2\hbar$~\cite{deFlorian:2014yva} (see Fig.~\ref{deltaG}), which is
nearly ten times smaller than what was originally expected~\cite{Altarelli:1988nr}. A recent summary of the RHIC spin achievements
can be found in Ref.~\cite{Aschenauer:2015eha}.

\section{Parton Orbital Motion and Discovery of Deeply Virtual Compton Scattering}

Although in the simple quark model, three quarks are in the $s$-wave state, there are
many evidences that the quarks do have substantial orbital motion. In the spin decomposition in the infinite momentum frame,
quark and gluon parton orbital angular momenta are parts of the contribution (see Eq.~(\ref{jaffe})). In 1996,
I came up with another way to decompose the nucleon spin~\cite{Ji:1996ek},
\begin{equation}
   {1\over2} = J_q+J_g = {1\over2}\Delta \Sigma + L_q^z + J_g\ ,
\end{equation}
where $J_g$ is the total gluon angular momentum and $L_q$ is the quark orbital angular momentum.
This decomposition is most natural in the rest frame of the nucleon. However,
it can be shown to be correct with a moving nucleon polarized either longitudinally or transversely.

The orbital angular momentum is generated through the orbital motion,
\begin{equation}
   \vec{L} = \vec{r}\times \vec{p}\ .
\end{equation}
To measure its contribution, one possibility is through the form factors of the field momentum density.
This situation is very similar to the magnetic moment which is a spatial moment of the electromagnetic
current. If one knows the form factors of the electromagnetic current,
\begin{equation}
\langle P'|j^\mu|P\rangle
= \bar U(P')\left[F_1(\Delta^2)\gamma^\mu + F_2(\Delta^2) {i\sigma^{\mu\nu} \Delta_\nu\over 2M}\right]U(P)\ ,
\end{equation}
where $\Delta= P'-P$, then the magnetic moment is just
\begin{equation}
\mu = (F_1(0)+F_2(0))\mu_N\ ,
\end{equation}
with $\mu_N$ being the nuclear magneton.

Thus, to measure the orbital angular momentum, we need to determine the matrix elements
of the quark and gluon momentum density,
\begin{equation}
    \langle P'|T^{0i}|P\rangle=\bar{U}(P') \left[A(\Delta^2) \gamma^{\{0} \bar{P}^{i\}} + B(\Delta^2) {\bar{P}^{\{0} i\sigma^{i\}\alpha} \Delta_\alpha\over2M} + C(\Delta^2){\Delta^0\Delta^i\over M}\right] U(P)\ ,
\end{equation}
where $T^{\mu\nu}$ is the energy-momentum tensor, $\bar{P}=(P+P')/2$. We need a process in which the proton is probed with
a current coupled with momentum density, and scatters elastically. However, the nature does not provide a direct momentum density probe.
Although we have discovered the gravitational wave, there is still a long way to go before we can use
gravity to study the internal structure of the nucleon.

\begin{figure}
\centering
\includegraphics[width=0.4\textwidth]{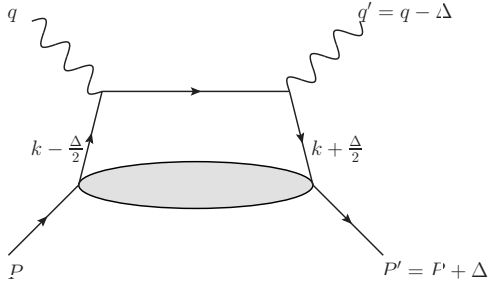}
\caption{Leading order Feynman diagram for the DVCS process.}
\label{dvcs}
\end{figure}

In late 1995, through theoretical arguments, I came up with a way to measure the form factors of the momentum density through a
process which I named as deeply virtual Compton scattering (DVCS). The initial kinematics of the DVCS is similar to DIS,
namely a high-energy lepton scattering off a proton target by exchanging a highly virtual,
high-energy photon. The final state, however, is totally different from those of DIS;
it consists of a high-energy real photon plus a slightly recoiled target proton, a highly exclusive process.
One would naively think the process is highly improbable; however, this is not the case, just like in high-energy
diffractive scattering. This process is in effect
a Compton scattering on a single quark, as shown in Fig.~\ref{dvcs}. In this kinematics, any alternative
scattering mechanism will be suppressed by $1/Q^2$, where $Q^2$ is the virtuality of the photon.
That this process probes the form factors of the momentum density can be seen from the
operator product expansion of the two electromagnetic currents,
\begin{equation}
              J^\mu(x)J^\nu(0)\sim C^{\mu\nu}_{~~\alpha\beta}(x) T^{\alpha\beta}(0) + \cdots\ ,
\end{equation}
which contains the quark and gluon parts of the energy-momentum tensor of QCD.

So I wrote a preprint and published it on arXiv in March 1996~\cite{Ji:1996ek}. In this paper, I calculated
the DVCS amplitude,
\begin{eqnarray}
T^{\mu\nu}(P,q,\Delta) &=& {1\over2}\left(g^{\mu\nu} - p^\mu n^\nu - p^\nu n^\mu\right)\int_{-1}^1 dx\left({1\over x-\xi/2+i\epsilon} + {1\over x+\xi/2-i\epsilon}\right)\nonumber\\
&&\times \left[H(x,\xi,\Delta^2) \bar{U}(P') \slashed n U(P) + E(x,\xi,\Delta^2) \bar{U}(P){i\sigma^{\alpha\beta} n_\alpha\Delta_\beta\over 2M} U(P)\right]\nonumber\\
&&+ {i\over2} \epsilon^{\mu\nu\alpha\beta}p_\alpha n_\beta \int_{-1}^1 dx \left({1\over x-\xi/2+i\epsilon} - {1\over x+\xi/2-i\epsilon}\right)\nonumber\\
&&\times \left[\tilde{H}(x,\xi,\Delta^2) \bar{U}(P') \slashed n \gamma_5 U(P) + E(x,\xi,\Delta^2){\Delta\cdot n\over 2M} \bar{U}(P)\gamma_5 U(P)\right]\ ,
\end{eqnarray}
where $q$ is the four momentum of the virtual photon, $p$ and $n$ are two light-like vectors with $p^\mu=\Lambda(1,0,0,1)$ and $n^\mu=(1,0,0,-1)/2\Lambda$. As one can see, it contains a whole new class of nucleon structure functions, $H$, $E$, $\tilde H$ and $\tilde E$ which
I called ``off-forward parton distributions" and later renamed as
``generalized parton distributions" (GPDs). The four distributions are hybrids of form factors
and parton distributions, depending on the Feynman parton momentum $x$, the t-channel momentum
transfer $\Delta^2 = t$, and the new kinematic variable $\xi= -n\cdot\Delta$ which is the projection
of the momentum transfer along the longitudinal direction. Once these new functions are known,
one can integrate to get the form factors of the momentum density,
\begin{equation}
                 \int^{+1}_{-1} dx\ x [H(x,\xi,t)+E(x,\xi,t)] = A(t) + B(t)\ ,
\end{equation}
while the total quark and gluon angular momentum is just
\begin{equation}
J_{q,g}= 1/2 [A_{q,g}(0)+B_{q,g}(0)] \ .
\end{equation}
I also stressed that DVCS belongs to a class of new Hard Exclusive Processes, in which the real photon can be
replaced by mesons, and these processes are equally useful in probing the GPDs.

I was very happy with these findings and decided to send the preprint to Physical Review Letters (PRL). After a couple of months,
I received responses. To my surprise, both referees rejected the paper and
questioned seriously the validity of my results. The first referee did not believe that the orbital angular
momentum can be measured. He/she wrote ``I do not understand why the author wants to construct spin and orbital angular momentum operators for the relativistic system where orbital angular momentum is an ill-defined concept. This is due to the fact that spin and orbital angular momentum can turn into each other by going to different Lorentz frames. They are well-defined and useful only for non-relativistic systems. Any definition in relativistic systems will involve some ad hoc assumption which might not be justified. It is not so clear to me how this is going to improve our understanding of the hadrons." The second referee believed that ``there is a flaw that makes this paper unsuitable for publication."
Specifically, he/she wrote ``Although the parton model diagram considered by the author supports this view, I believe that this would not survive radiative corrections. In fact, off-forward amplitudes for scattering of colored objects are not infrared finite (the same thing happens in electromagnetism). The examination of one soft gluon correction to the processed considered by the author already shows the IR divergence." Based on these, the editor rejected the paper.

I was not totally upset by these negative reports because I believed both objections can be overcome. So I wrote a long and detailed reply to both referees,
pointing out that the orbital angular momentum is meaningful as long as there is a transverse momentum, and the transverse components of a four-vector are not affected by Lorentz boosts etc. I argued that although I do not have a detailed proof at hand, the process should be infrared factorizable---namely, all soft divergences cancel and collinear divergences can be absorbed into the new parton distributions---by pointing out a few relevant signs including a one-loop calculation. I even quoted my private communication
with a world-class expert in factorization, John Collins, who believes this process is factorizable.

I was a bit frustrated when I saw the second round of reports another two months later. The first referee started the report with
``I stand by my previous report." He/she went on saying ``in the direction of motion,
there is no orbital angular momentum, because its projection is zero." I finally understood that the referee was confused about the orbital angular momentum
between hadron and the underlying parton. The second referee insisted that ``there is a serious problem" with the infrared physics, and ``no credible answer is given in the paper as it stands", namely, he/she wanted to see an all-order proof of factorization, which was not in the paper.

I decided to appeal to the Divisional Associate Editors, as I believe the paper is important and correct.
In September, I got a reply, in which the first paragraph read, ``I have studied the file
relating to the paper `Gauge-Invariant Decomposition of Nucleon Spin' by X. Ji, and have concluded that it should be published in Physical Review Letters. The paper was originally submitted in March. Already by the end of April, a paper submitted to Physics Letters B by no less an investigator than A. Radyushkin of the Jefferson Laboratory, which has since appeared, which was inspired by this paper, and which cites it as the first reference in the first sentence. This by itself argues strongly for publication without further delay." In the end of the reply, signed George Sterman. I was very grateful for this decision. Later when I asked George why he was sympathetic with my appeal, he told me a similar story with his ``jet" paper where the all-important idea of hadron jet was first introduced~\cite{Sterman:1977wj}. Even with co-author Steven Weinberg did not change the fate of repeated
rejections from PRL referees.

The paper eventually appeared in PRL at the end of January, 1997, and the term DVCS was thus officially in print. The paper now has over 1400 citations
according to HEP data base (https://inspirehep.net/). I followed up the letter paper with a longer version,
titled ``Deeply Virtual Compton Scattering" in which I presented the detailed derivation of DVCS cross section and its interference with the Bethe-Heitler process~\cite{Ji:1996nm}. I also presented the one-loop evolution equations
for GPDs which interpolated between Dokshitzer-Gribov-Lipatov-Altarelli-Parisi€"€"€"€" evolution for parton distributions and the Efremov-Radyushkin-Brodsky-Lepage evolution kernel for meson wave function amplitudes.

\begin{figure}
\centering
\includegraphics[width=0.5\textwidth]{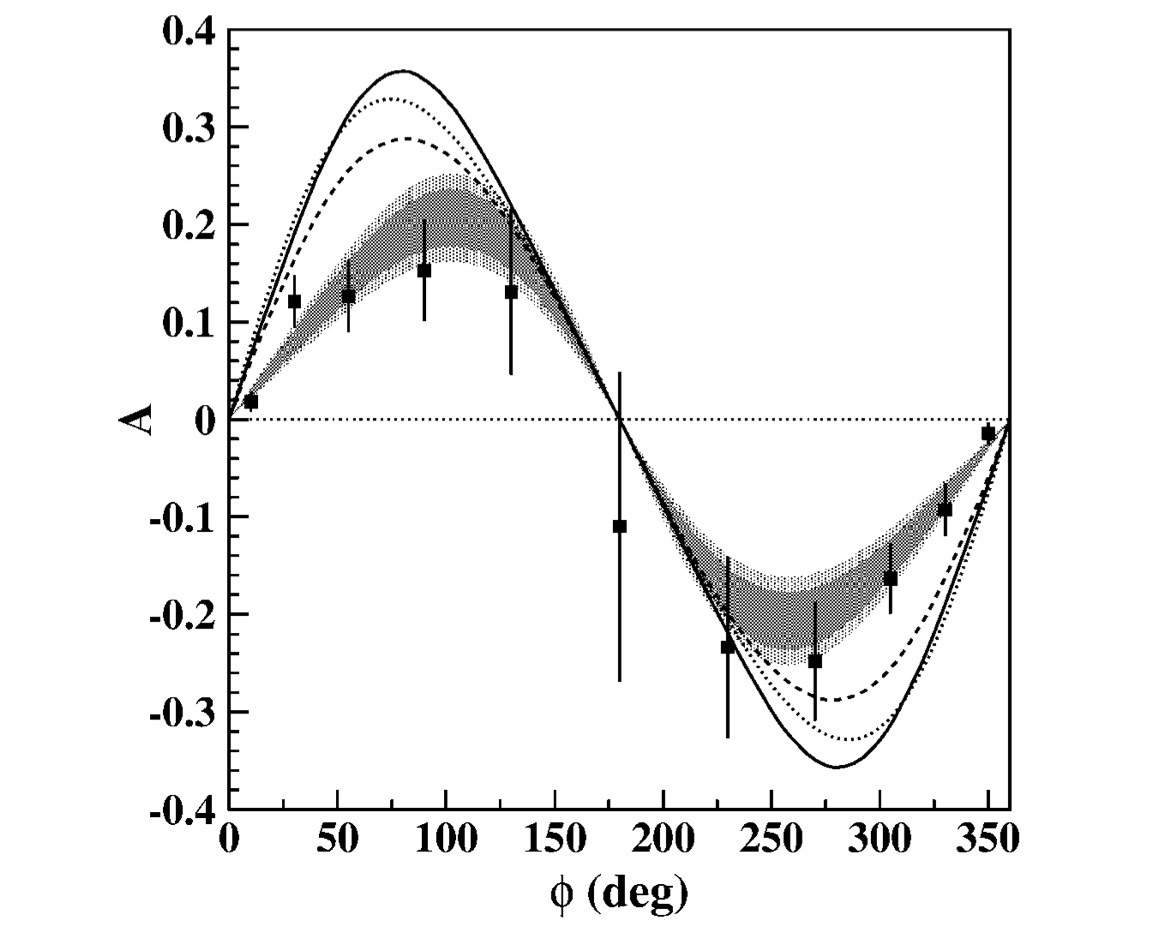}
\caption{Azimuthal angle $\phi$ dependence of the beam-spin asymmetry~\cite{Stepanyan:2001sm}. The $\sin\phi$ modulation is predicted by the DVCS theory.}
\label{clas}
\end{figure}

The experimental observation of the DVCS process came in 5 years after the theoretical suggestion of its existence.
In June 2001, HERMES collaboration at DESY published ``Measurement of the Beam-Spin Azimuthal Asymmetry Associated with Deeply-Virtual Compton Scattering"~\cite{Airapetian:2001yk}. In July, H1 collaboration at DESY published ``Measurement of Deeply Virtual
Compton Scattering at HERA"~\cite{Adloff:2001cn}; CLAS collaboration at Jefferson Lab published ``Observation of Exclusive Deeply Virtual Compton Scattering in Polarized Beam Asymmetry Measurement"~\cite{Stepanyan:2001sm}. Fig.~\ref{clas} shows the azimuthal angle dependence of the beam-spin asymmetry measured by CLAS collaboration, which indicates the existence of DVCS effects. By now, COMPASS at CERN, H1, ZEUS and HERMES at DESY, and Hall A and B at Jefferson Lab all have
made many measurements of this process and other deep-exclusive processes with different combinations of beam and target polarizations on proton,
deuteron and helium-3 targets. A search of Inspire database showed 411 papers with the word ``deeply virtual" in titles.

\begin{figure}
\centering
\includegraphics[width=0.6\textwidth]{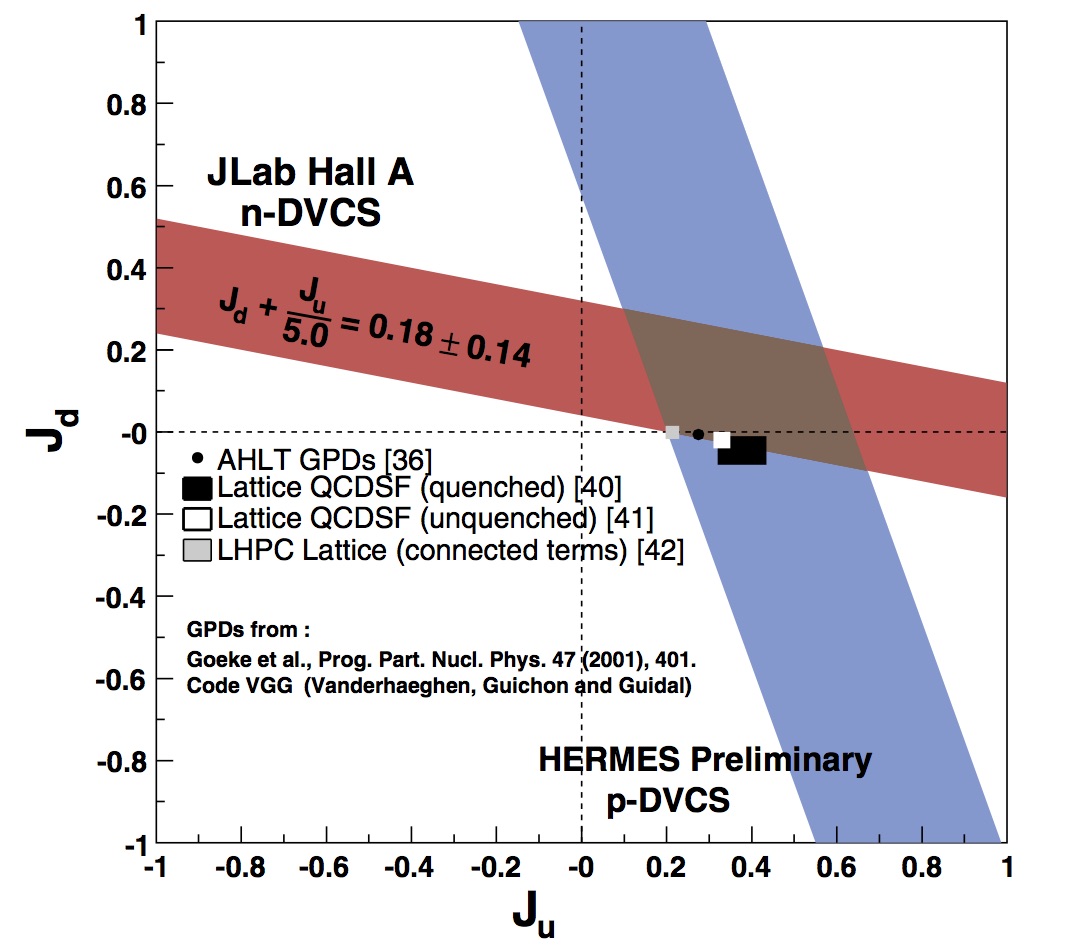}
\caption{Constraints on quark angular momentum from experimental extractions and lattice QCD calculations~\cite{Mazouz:2007aa}.}
\label{jlab}
\end{figure}

We shall have a program on the systematic analysis of experimental data. Some levels of modeling on GPDs are needed just like in the case of ordinary parton distributions. However, model dependencies must be studied carefully and reduced to a minimal level. What one hopes to get in the end is, among others, the
angular momentum contributions to the spin of the proton from individual quark flavors (see Fig.~\ref{jlab}~\cite{Mazouz:2007aa}).
Meanwhile, one can also make precision lattice QCD calculations, where the main uncertainty
comes from quark loop contributions.

It is proper to acknowledge the other relevant contributions. A group at Leipzig recognized the existence of
GPDs a couple of years before my work, which was published in an East-Germany magazine~\cite{Mueller:1998fv}.
Although they considered theoretically-possible two-virtual-photon process, they did not consider the practical DVCS
with real photon final state as in my work (hence there is no infrared question as discussed above), nor
did they realize any connection to the spin structure of the nucleon and orbital angular momentum.
I was informed of their work when attending the international SPIN conference in Amsterdam in September, 1996.
In the published version of my PRL paper, a reference was added to acknowledge that GPDs and their scale evolutions had been
studied earlier. In spring 1996, an INT (Institute for Nuclear Theory) program was organized on the quark and gluon structure of nucleons and nuclei from Feb. 20 to May 31. A. Radyushkin and I were participants. After seeing my preprint, he believed that the DVCS process I suggested should be
described by what he called double distributions. He quickly
wrote a paper on the scaling limit of DVCS with double distributions~\cite{Radyushkin:1996nd} and got it published.
Although he did not realize at that time, double distributions and his alternative analysis of DVCS are
entirely equivalent to what I had done earlier. However, double distributions
do provide an alternative way to parameterize GPDs. In the following years,
Anatoly played a key role in promoting theoretical and experimental studies of DVCS and double distributions.

\section{Phase-Space Tomography and JLab 12 GeV upgrade, Electron-Ion Collider}

What do the GPDs tell us about the structure of the nucleon besides the orbital angular momentum? The answer is that
it provides a detailed map of the quarks and gluons in the nucleon interior
in phase space or a phase-space tomography. This is very exciting
because the traditional approach of elastic and deep-inelastic scattering on the nucleon provides
only the static coordinate or momentum space pictures, separately. The GPDs provide pictures of
dynamical correlations in both coordinate and momentum spaces.

Historically, every discovery of a new probe has generated major
advances in understanding the nucleon structure. In the 1950's, Hofstadter and collaborators made
systematic studies of elastic electron scattering off the nucleons and measured the elastic form factors
of electromagnetic currents for the first time~\cite{Hofstadter:1956qs}. They provided important information of spatial charge and magnetization
distributions of the nucleon. These distributions are particularly easy to understand in the so-called impact parameter
space of two transverse dimensions when the nucleon is moving at the speed of light~\cite{Miller:2007uy}. In the late 1960's
and early 70's, Friedman, Kendall and Taylor made pioneering studies of DIS
of electrons off the nucleons, discovered quarks and measured their longitudinal momentum distributions
in a fast moving nucleon~\cite{Friedman:1972sy}. Today, parton distributions are essential tools for studying
new physics at colliders~\cite{Nadolsky:2008zw}.

\begin{figure}
\centering
\includegraphics[width=0.7\textwidth]{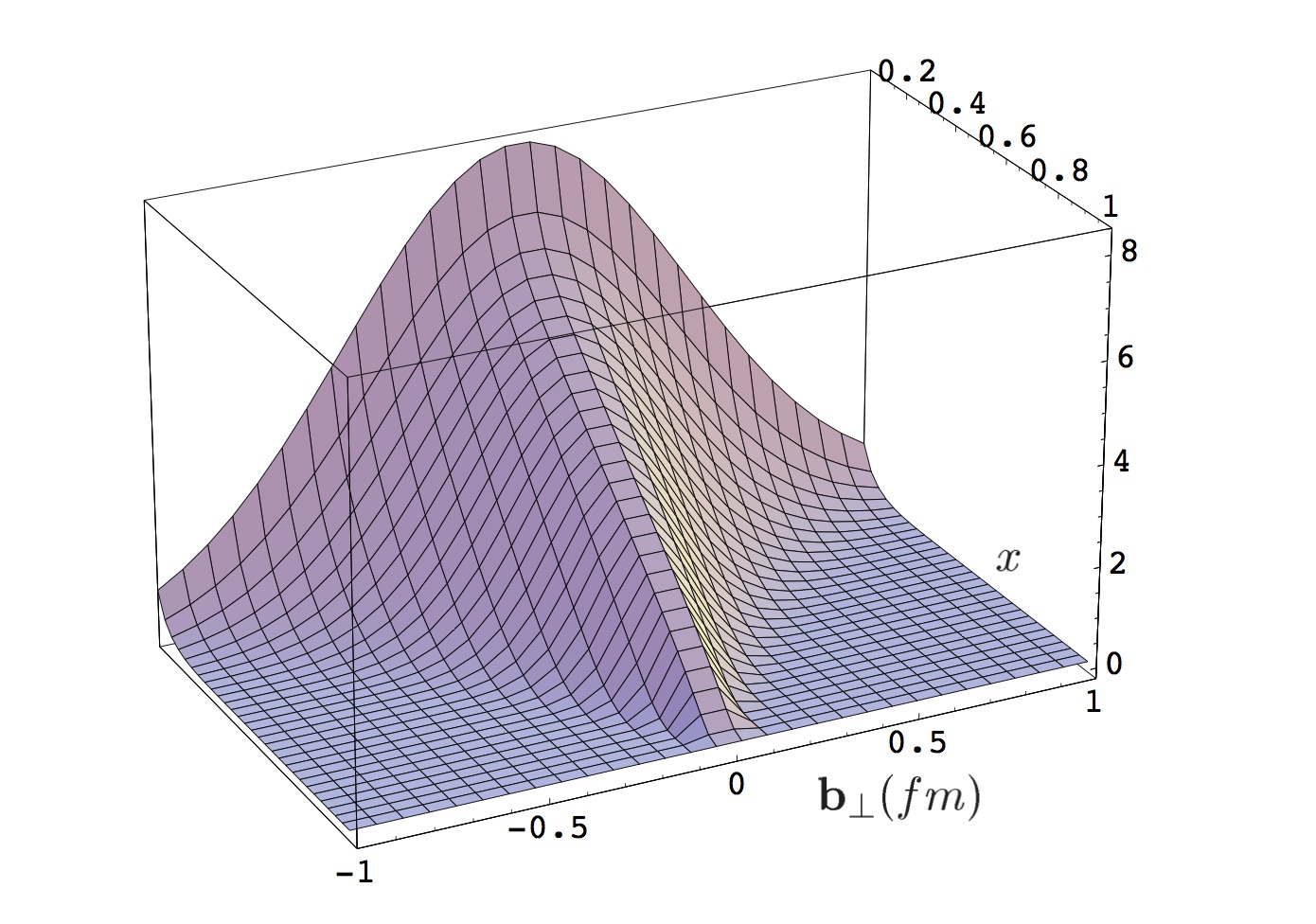}
\caption{Impact parameter dependent $u(x,\vec{b}_\perp)$ for a simple model~\cite{Burkardt:2002hr}.}
\label{gpd}
\end{figure}

A natural extension of these well-studied nucleon observables would be correlated distributions
in both momentum and coordinate spaces. One such quantity is parton distributions in the impact parameter
space. However, for a long time, nobody knows how to connect it to a physical observable.
In 2000, Matthias Burkardt
published a paper which showed that the Fourier transform of a special projection of GPDs at $\xi=0$
gives the images of partons in the impact parameter space~\cite{Burkardt:2000za}. This is an important physical
interpretation of the GPDs. Notice the hybrid nature of this impact-parameter space parton density.
Since one probes the transverse coordinates and longitudinal momentum in separate dimensions, quantum mechanical
uncertainty principle is not an issue here. Fig.~\ref{gpd} shows the impact-parameter-dependent $u$ quark
distribution for a simple model~\cite{Burkardt:2002hr}.

\begin{figure}
\centering
\includegraphics[width=\textwidth]{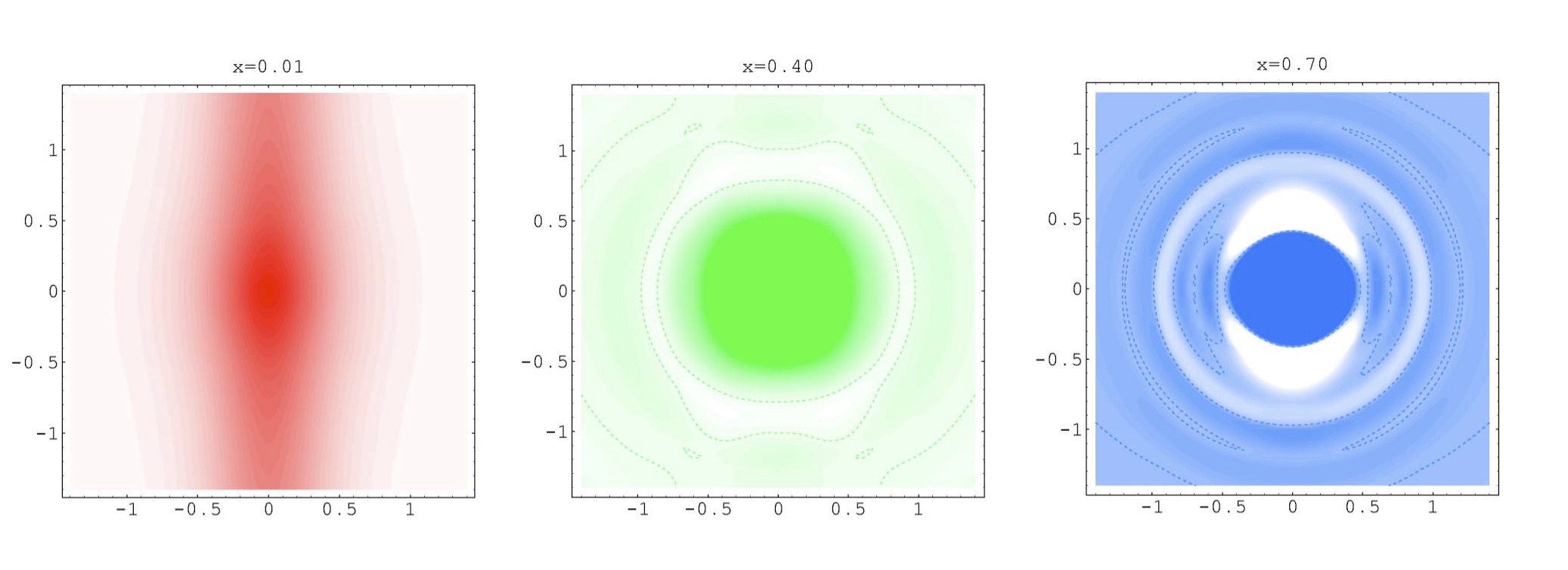}
\caption{The $u$-quark phase-space charge distribution at different values of $x$ for an {\it ansatz} of GPD~\cite{Belitsky:2003nz}. The vertical and horizontal axes are $z$ and $|\vec{r}|$, respectively, measured in femtometers. The (dashed) contours separate regions of positive (darker areas) and negative (lighter areas) densities.}
\label{quarkwigner}
\end{figure}

Belitsky, Yuan and I extended Burkardt's work to the full GPD by introducing the phase-space
Wigner distribution~\cite{Belitsky:2003nz}, which is a quantum mechanical analog of the classical phase space
distribution in statistical mechanics. It is a 6-dimensional distribution that involves both
3-coordinate and 3-momentum. After integrating over the transverse momentum, one is left with a 4-dimensional object, whose
Fourier transform is related to the full GPDs. This 4-dimensional distribution
allows us to construct 3D images of quarks for very fixed longitudinal momentum (see Fig.~\ref{quarkwigner}). Thus
a full measurement of GPDs will allow a detailed phase-space
tomography of the nucleons. A more recent application of these 6D Wigner distributions has been a reduction 
to 5-dimensional objects~\cite{Lorce:2011kd} which has provided new insights on differences between various definitions 
of quark orbital angular momentum. The 5D Wigner distribution might measurable at least in the small-$x$ limit~\cite{Hatta:2016dxp}.

The Jefferson Lab 12 GeV upgrade has been nearly finished, and an exciting physics program
is about to begin~\cite{12gev}. Jefferson Lab 12 GeV facility with enhanced capabilities in existing halls and increased
luminosity of $10^{35}$ to $10^{39} /cm^2 s$ will set an important milestone for measuring DVCS
and related hard exclusive processes. They are extensively planned in all three
existing experimental halls~\cite{angela}. In Hall A, there is the E12-06-114 experiment aiming at measuring
the absolute cross sections and test of scaling in $Q^2$ with extended kinematic coverage. This will
be the first experiment to run after 12 GeV upgrade. The CLAS12 DVCS experiments will have
large kinematic coverage, measure various types of spin asymmetries, and perform the
Compton form factor extractions. Hall C E12-13-010 experiment will perform energy separation
of the DVCS cross section, measure low $x_B$ as well as high $Q^2$, and study higher-twist
contributions.

However, an ideal machine to measure hard exclusive processes will need: higher energy so that
the contributions from simple parton scattering mechanism dominate, high luminosity to get
good enough statistics, and capability of detecting exclusive events with high resolutions of event topology.
This calls for an Electron-Ion Collider (EIC).

\begin{figure}
\centering
\includegraphics[width=0.8\textwidth]{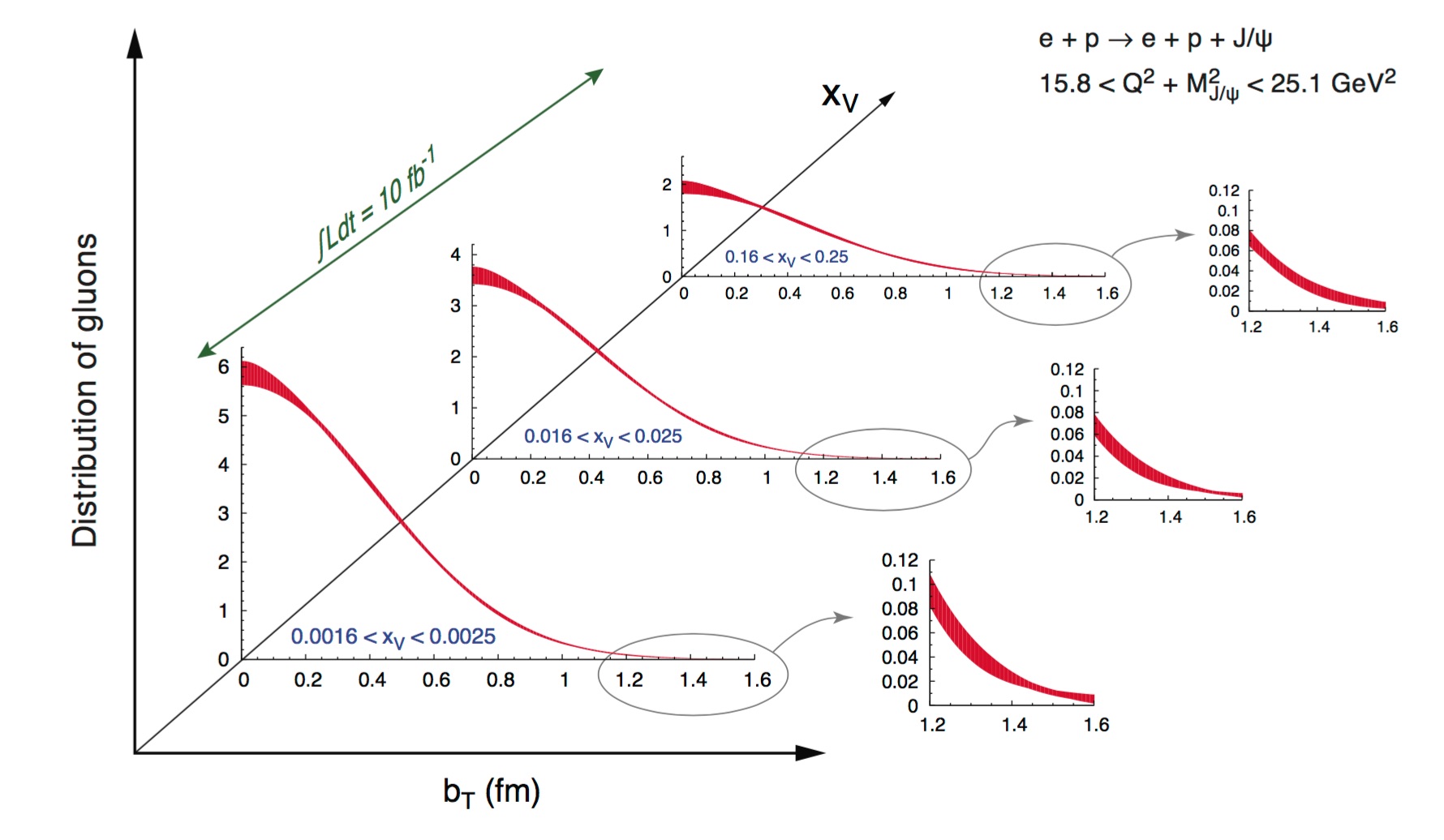}
\caption{The projected precision of the transverse spatial distribution of gluons as obtained from the cross section of exclusive $J/\Psi$ production~\cite{Accardi:2012qut}.}
\label{gluongpd}
\end{figure}

The nuclear science community in the United States have published the 2015 long-range plan for nuclear
science~\cite{nsac}, in which the recommendation 3 is specially about a future EIC in the United States. It states ``We recommend
a high-energy high-luminosity polarized Electron-Ion Collider as the highest priority for new facility
construction after the completion of FRIB", and it will be used for ``precise imaging of gluons
in the nucleons and nuclei", studying ``the origin of nucleon spin", etc. An updated white paper
for EIC is available online now~\cite{Accardi:2012qut}.
If the Jefferson Lab 12 GeV facility provides precise phase-space images of the quarks in the large $x$ region,
then EIC will provide similar images at small $x$ and for gluons. Shown in Fig.~\ref{gluongpd} is the projected precision of the transverse spatial distribution of gluons as obtained from the cross section of hard-exclusive $J/\Psi$ production~\cite{Accardi:2012qut}.

\vspace{.5cm}

To summarize, the deeply virtual compton process was first discovered in theory 20 years ago. At DESY, Jefferson Lab 6 GeV,
CERN COMPASS, we have already seen the initial physical potential of DVCS and related hard exclusive processes
in probing the internal structure of the protons and neutrons. At Jefferson Lab 12 GeV, we expect to learn
a lot more about DVCS; this is extremely exciting and there is a lot to do in theory and experiment
in the next few years to bring out the detailed images of quarks in the nucleons. Ultimately, an EIC with high luminosity
and dedicated exclusive-event detectors will be the ideal place for a complete phase-space tomography of the proton.

\vspace{1.0cm}

I thank my Ph.~D. and postdoctoral advisors, H. Wildenthal, S. Koonin, J. Negele, and R. Jaffe for bringing
and keeping me in the field, my theory friends, S. Brodsky, M. Burkardt, J. Collins, L. Gamberg, P. W-Y. Hwang, X. G. He, H. N. Li, K. F. Liu, M. Ramsey-Musolf,
G. Miller, A. Mueller, J. W. Qiu, A. Radyushkin, A. Schaefer, G. Sterman, A. W. Thomas,
W. Vogelsang, etc. for showing me that theoretical physics is fun, and experimental friends H. Avakian, D. Beck, V. Burkert, G. Cates, J. P. Chen, L. Elouadrhiri, H. Gao, D. Hertzog, B. Filippone,
E. Hughes, N. Makins, R. McKeown, Z. Meziani, R. Milner, A. Nathan, J. C. Peng, M. Perdekamp, N. Saito, etc,  for keeping an interest in my theoretical work, and last but not the least my collaborators, postsocs and students, I. Balitsky, A. Belitsky, J. W. Chen, W. Melnitchouk, J. Osborne, F. Yuan, J. H. Zhang, and Y. Zhao, etc,  for
fruitful collaborations. My research has been supported by the U. S. Department of Energy
and National Science Foundation of China. Y. Zhao's assistance in finishing this write-up is appreciated.

\end{document}